\documentclass[aps,pre,floats,twocolumn,showpacs,superscriptaddress]{revtex4}

\usepackage{graphicx,epsfig}
\usepackage{times}
\usepackage{graphics,dcolumn,bm,fleqn,float}
\usepackage{amssymb,amsmath,multirow,rotate,color}
\begin{document}

\title{Co-evolutionnary network approach to cultural dynamics controlled by intolerance.}

\author{Carlos Gracia-L\'azaro}

\affiliation{Institute for Biocomputation and Physics of Complex
Systems (BIFI), University of Zaragoza, Zaragoza 50018, Spain}

\affiliation{Departamento de F\'{\i}sica de la Materia Condensada,
University of Zaragoza, Zaragoza E-50009, Spain}

\author{Fernando Quijandr\'{\i}a}

\affiliation{Laboratoire de Physique Th\'eorique et Mod\'elisation, UMR CNRS. Universit\'e de Cergy-Pontoise, 
2 Avenue Adolphe Chauvin, 95302, Cergy-Pontoise Cedex, France}

\author{Laura Hern\'andez}

\affiliation{Laboratoire de Physique Th\'eorique et Mod\'elisation, UMR CNRS. Universit\'e de Cergy-Pontoise, 
2 Avenue Adolphe Chauvin, 95302, Cergy-Pontoise Cedex, France}

\author{Luis Mario Flor\'{\i}a}

\affiliation{Institute for Biocomputation and Physics of Complex
Systems (BIFI), University of Zaragoza, Zaragoza 50018, Spain}

\affiliation{Departamento de F\'{\i}sica de la Materia Condensada,
University of Zaragoza, Zaragoza E-50009, Spain}

\author{Yamir Moreno}

\affiliation{Institute for Biocomputation and Physics of Complex
Systems (BIFI), University of Zaragoza, Zaragoza 50018, Spain}

\affiliation{Departamento de F\'{\i}sica Te\'orica. University of
Zaragoza, Zaragoza E-50009, Spain}

\affiliation{Complex Networks and Systems Lagrange Lab, Institute for
   Scientific Interchange, Viale S. Severo 65, 10133 Torino, Italy}

\date{\today}

\begin{abstract}

Starting from Axelrod's model of cultural dissemination, we introduce a rewiring probability, enabling
agents to cut the links with their unfriendly neighbors if their cultural similarity is below a tolerance
parameter. For low values of tolerance, rewiring promotes the convergence to a frozen monocultural state. However,
intermediate tolerance values prevent rewiring once the network is fragmented, resulting in a multicultural society
even for values of initial cultural diversity in which the original Axelrod model reaches globalization.
\end{abstract}


\maketitle

\section{Introduction}
\label{I}

The growing interest in the interdisciplinary physics of complex systems, 
has focussed physicists' attention on agent-based modeling \cite{Axel1,Epstein} 
of social dynamics, as a very attractive methodological framework for social 
sciences where concepts and tools from statistical physics turn out to be 
very appropriate \cite{castellano2} for the analysis of the collective behaviors 
emerging from the social interactions of the agents. The dynamical social 
phenomena of interest include residential segregation \cite{Sch1,Sch2}, 
cultural globalization \cite{Axel2,castellano1}, opinion formation \cite{Galam,15m}, rumor 
spreading \cite{Daley,mnv} and others.

The question that motivates the formulation of Axelrod's model for cultural 
dissemination \cite{Axel2} is how cultural diversity among groups and individuals 
could survive despite the tendencies to become more and more alike as a result of social 
interactions. The model assumes a highly non-biased scenario, where the culture 
of an agent is defined as a set of equally important cultural features, whose particular 
values (traits) can be transmitted (by imitation) among interacting agents. It also 
assumes that the driving force of cultural dynamics is the "homophile satisfaction", 
the agents' commitment to become more similar to their neighbors. Moreover, 
the more cultural features an agent shares with a neighbor, the more likely the 
agent will imitate an uncommon feature's trait of the neighbor agent. In other words, 
the higher the cultural similarity, the higher the social influence.   

The simulations of the model dynamics show that for low initial cultural diversity, 
measured by the number $q$ of different traits for each cultural feature (see below), 
the system converges to a global cultural state, while for $q$ above a critical 
value $q_c$ the system freezes in an absorbing state where different cultures 
persist. The (non-equilibrium) phase transition \cite{Marro} between globalization 
and multiculturalism was first studied for a square planar geometry 
\cite{castellano1,vilone,vazquez1}, but soon other network structures of social 
links \cite{Klemm1,Klemm1a,guerra} were considered, as well as the effects of 
different types of noise ("cultural drift") \cite{Klemm2,Klemm2a}, 
external fields (modeling {\em e.g.} influential media, or information feedback)
\cite{Shibanai,Gonzalez1,Gonzalez3,rodriguez}, and global or local non-uniform 
couplings \cite{Gonzalez4,Gonzalez2}. 

In all those extensions of Axelrod's model mentioned in the above paragraph, 
the cultural dynamics occurs on a network of social contacts that is fixed from the 
outset. However, very often social networks are dynamical structures that continuously 
reshape. A simple mechanism of network reshaping is agents' mobility, and a scenario 
(named the Axelrod-Schelling model) where cultural agents placed in culturally dissimilar 
environments are allowed to move has recently been analyzed \cite{Ax_Sch1,Ax_Sch2}. In this 
model, new interesting features of cultural evolution appear depending on the values of 
a parameter, the (in-)tolerance, that controls the strength of agents' mobility. 

A different mechanism of network reshaping has been considered in \cite{vazquez2,centola}, 
where a cultural agent breaks its link to a completely dissimilar neighbor, redirecting 
it to a randomly chosen agent. At variance with the mobility scenario of the Axelrod-Schelling 
model, that limits the scope of network structures to clusters' configurations on the starting 
structure (square planar lattice, or others), the rewiring mechanism allows for a wider set of 
network structures to emerge in the co-evolution of culture and social ties. 

In this paper we introduce in the scenario of network rewiring a tolerance parameter $Z$ 
controlling the likelihood of links rewiring, in such a way that the limit $Z=1^-$ recovers the 
case analyzed in \cite{vazquez2,centola}, where only links with an associated null cultural 
overlap are broken. Lower values of $Z$ correspond to less tolerant attitudes where social links 
with progressively higher values of the cultural overlap may be broken with some probability 
that depends on these values. The results show a counterintuitive dependence of the tolerance $Z$ on the critical value $q_c$. On one hand, as 
expected from \cite{vazquez2,centola}, rewiring promotes globalization for high values of the tolerance, but on the other hand, very low
values of $Z$ (which enhance the rewiring  probability) show the higher values of $q_c$. Indeed, a non monotonous 
behavior is observed in $q_c(Z)$: Our results unambiguously show that for some intermediate
values of the tolerance $Z$, cultural globalization is disfavored with respect to the original Axelrod's model where 
no rewiring of links is allowed. In other words, rewiring does 
not always promote globalization. On the other hand, the resulting network topology depends 
on $q$, changing from a Poisson connectivity distribution $P(k)$ to a fat tailed 
distribution for $q\sim q_c$.

\section{The model}
\label{II}

As in Axelrod's model, the culture of an agent $i$ is a vector of $F$ integer variables 
$\{\sigma_f(i)\}$ ($f=1,...,F$), called cultural {\em features}, that can take on $q$ 
values, $\sigma_f(i) = 0,1,...,q-1$, the cultural {\em traits} that the feature $f$ can assume. 
The $N$ cultural agents occupy the nodes of a network of average degree $\langle k \rangle$ 
whose links define the social contacts among them. The dynamics is defined, at each time step, 
as follows:

\begin{itemize}

\item Each agent $i$ imitates an uncommon feature's trait of a randomly chosen 
neighbor $j$ with a probability equal to their {\em cultural overlap} $\omega_{ij}$, 
defined as the proportion of common cultural features, 
\begin{equation}
\omega_{ij} =
\frac{1}{F}\sum_{f=1}^{F}\delta_{\sigma_f(i),\sigma_f(j)},
\label{overlap}
\end{equation}
where $\delta_{x,y}$ denotes the Kronecker's delta which is 1 if $x=y$
and 0 otherwise. The whole set of $N$ agents perform this step in parallel.

\item Each agent $i$ disconnects its link with a randomly chosen neighbor 
agent $j$ with probability equal to its {\em dissimilarity} $1-\omega_{ij}$, 
provided the dissimilarity $1- \omega_{ij}$ exceeds a threshold ({\em tolerance}) $Z$, 
\begin{equation}
1- \omega_{ij} > Z\;,
\label{rewiring}
\end{equation}
and rewires it randomly to other non-neighbor agent. The tolerance $0 \leq Z \leq 1$ 
is a model parameter.

\end{itemize}

First we note that the initial total number of links in the network is preserved 
in the rewiring process, so the average degree $\langle k \rangle$ remains constant.
However, the rewiring process allows for substantial modifications of the network 
topological features, {\em e.g.} connectedness, degree distribution, etc. In that respect,
except for the limiting situation of very low initial cultural diversity $q$ and a very high 
tolerance $Z$ (where the likelihood of rewiring could be very low), one should expect 
that the choices for the initial network of social ties have no influence in the asymptotic 
behavior of the dynamics.
  
When the threshold tolerance $Z$ satisfies $\frac{F-1}{F} \leq Z <1$, only those 
links among agents with zero cultural overlap are rewired, so the model becomes the 
one studied in \cite{vazquez2,centola}. On the other hand, when the tolerance takes
the value $Z=1$, there is not rewiring likelihood and the original Axelrod's model is recovered. 
When $Z=0$ rewiring is always possible provided the cultural similarity is not complete, 
{\em i.e.,} $\omega_{ij} \neq 1$, so that it corresponds to the highest intolerance.

The usual order parameter for Axelrod's model is  
$\langle S_{max} \rangle/N$, where $\langle S_{max}\rangle$ 
is the average (over a large number of different random initial conditions) 
of the number of agents sharing the most abundant (dominant) culture, 
and $N$ is the number of agents in the population. Large values of the order
parameter characterize the globalization (cultural consensus) regime. We also 
compute the normalized size $\langle S_{top} \rangle/N$ of the largest network 
component ({\em i.e.}, the largest connected subgraph of the network).

\section{Results and discussion}
\label{III}

We have studied networks of sizes $N=900$, $1600$; averaging over $50$ - $2000$ replicas. 
The considered cultural vectors have $F=10$ cultural features, each one with a variability 
$q=$ $5$ - $10000$. We studied different values of the tolerance threshold $Z \in (0,1)$ and 
different values of the average connectivity $\langle k \rangle= 4,10,20,40$.  Each simulation 
is performed for $N$, $F$, $\langle k \rangle$, $Z$, and $q$ fixed. For the sake of comparison 
with previous results~\cite{vazquez2,centola}, we will present results for $\langle k \rangle=4$.

\begin{figure}
\begin{center}
\epsfig{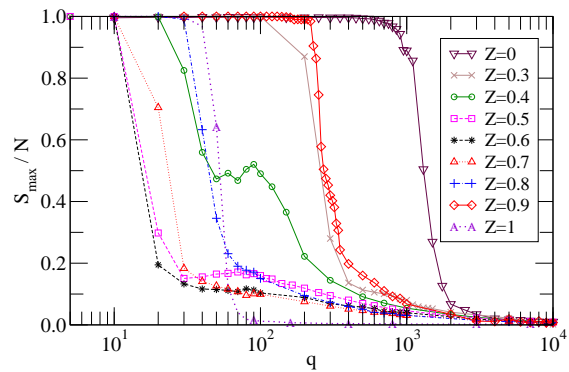}
\end{center}
\caption{(color online) Order parameter as a function of the variability $q$ for different values of the tolerance
threshold $Z$. $N=900$, $\langle k \rangle=4$, average over $1000$ replicas. }
\label{k4_diff_z}
\end{figure}

The behavior of the order parameter for different values of $Z$ is seen in Fig.~\ref{k4_diff_z}. 
Like in \cite{vazquez2}, three different macroscopic phases are observed with increasing values 
of $q$, namely a monocultural phase, with a giant cultural cluster, a multicultural one with 
disconnected monocultural domains, and finally a multicultural phase with continuous rewiring. 
The nature of the latter phase has been successfully explained in \cite{vazquez2}: At very large 
values of the initial cultural diversity $q$, the expected number of pairs of agents sharing at least 
one cultural trait becomes smaller than the total number of links in the network, so that rewiring 
cannot stop. Here we will focus attention on the first two phases and the transition between them. 

\begin{figure}
\begin{center}
\epsfig{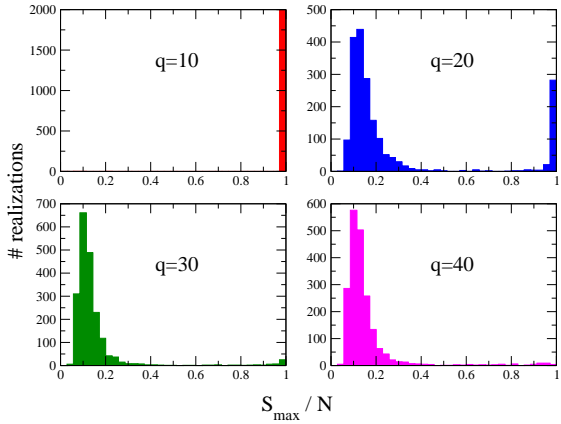}
\end{center}
\caption{(color online) Histograms of $S_{max}/N$ for different values of $q$, and for a fixed tolerance 
$Z=0.5$, $N=900$, $\langle k \rangle=4$. From this figure one gets $q_c \sim 20$.
 }

\label{histog}
\end{figure}

\begin{figure}
\begin{center}
\epsfig{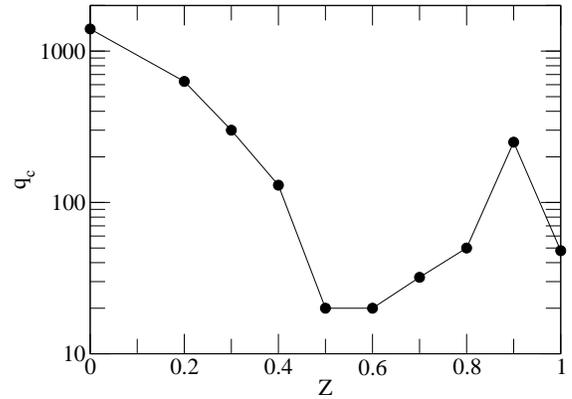}
\end{center}
\caption{ Critical value of the diversity $q_c$ versus the tolerance threshold $Z$, obtained from
the distribution of sizes of the dominant culture. $N=900$, $\langle k \rangle=4$. See the text for further details.}
\label{qc_vs_z}
\end{figure}

In figure~\ref{histog} we show the size distribution of the dominant culture over different realizations, 
measured for different values of $q$, at a particular fixed value of the tolerance $Z=0.5$. In the region 
of $q$ values near the transition from globalization to multiculturalism, the distribution is double peaked, 
indicating that the transition is first order, as in the original Axelrod's model. The transition value, $q_c$, 
may be roughly estimated as the $q$ value where the peaks of the size distribution are equal in height. 
The estimates of the transition points for different values of the tolerance $Z$ are shown in 
Fig.~\ref{qc_vs_z}. The non monotonous character of the graph $q_c(Z)$ seen in this figure reveals a 
highly non trivial influence of the tolerance parameter on the co-evolution of cultural dynamics and 
the network of social ties.  

Let us first consider the (most tolerant) case $Z=0.9$ that, except for the system size $N$, 
coincides exactly with the situation considered in \cite{centola}, {\em i.e.}, only links with zero 
cultural overlap are rewired. As discussed in \cite{centola}, for $q$ values larger than the critical 
value for a fixed network ( $q_c(Z=1) \simeq 60$), rewiring allows redirecting links with zero 
overlap to agents with some common cultural trait (compatible agents), so reinforcing the power 
of social influence to reach cultural globalization.  Once all links connect compatible agents, 
rewiring stops \cite{note}. From there on, the network structure will remain fixed, 
and globalization will be reached with the proviso that the network has so far remained connected. 
This is the case for most realizations (for $N=900$) up to values of $q \sim 240$. Increasing 
further the cultural diversity $q$, increases the frequency of rewiring events and slows down 
the finding of compatible agents, favoring the topological fragmentation into network components 
before rewiring stops. Under these conditions, the asymptotic state will consist of disconnected 
monocultural components. 

On one hand, network plasticity allows to connect compatible agents, so promoting globalization; 
but on the other hand it may produce network fragmentation, so favoring multiculturalism. What we 
have seen in the previous paragraph is that for $Z=0.9$ the first effect prevails over the second one up 
to $q_c(Z=0.9) \simeq 240$. Going from there to less tolerant situations (decreasing $Z$), increases 
the likelihood of rewiring, making easier that network fragmentation occurs before rewiring stops. 
This has the effect of decreasing the critical value $q_c$. In fact, from Fig.~\ref{qc_vs_z} we see 
that for $Z= 0.7, 0.6,$ and $0.5$ multiculturalism prevails for cultural diversities where the original Axelrod's 
model shows cultural globalization. In these cases network plasticity promotes multiculturalism in a 
very efficient way: Agents segregate from neighbors with low cultural similarity and form disconnected 
social groups where full local cultural consensus is easily achieved, for $q$ values low enough to allow a global culture in fixed connected networks.

\begin{figure}
\begin{center}
\epsfig{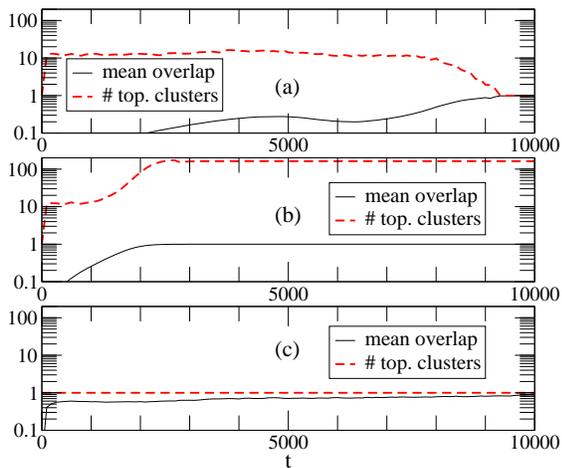}
\end{center}
\caption{(color online) Time evolution of mean overlap and number of topological clusters for diferent values of
tolerance $Z=0.2$(a), $Z=0.6$(b), $Z=0.9$(c). $N=900$, $q=100$. See the text for further details.}
\label{time.evolution}
\end{figure}

For very low values of the tolerance parameter, though network fragmentation occurs easily during the evolution, 
Fig.~\ref{qc_vs_z} shows that globalization persists up to very high values of the initial cultural 
diversity $q$. To explain this seemingly paradoxical observation, one must realize that network 
fragmentation is not an irreversible process, provided links connecting agents with high cultural 
overlap have a positive rewiring probability. Under these circumstances, transient connections among different components occur so frequently so as to make 
it possible a progressive cultural homogenization between components that otherwise would 
have separately reached different local consensuses. Fig.~\ref{time.evolution} illustrates the time evolution for $q=100$ and different values of $Z$. Panel (a) shows an example of cultural evolution where network fragmentation reverts to a connected 
monocultural network for $Z=0.2$. Panel (b), that corresponds to $Z=0.6$, shows that social 
fragmentation persists during the whole evolution, while in panel (c), which corresponds to the most tolerant situation (
$Z=0.9$), the network remains connected all the time.

The degree distribution of the network is Poissonian centered about $\langle k \rangle$ for all 
$q$ values, except for $q \gtrsim q_c$  where it becomes fat tailed, with several lowly connected 
(and disconnected) sites. For very high $q$ values, in the dynamical phase, the network rewiring 
is esentially random, so $P_q(k)$ is again Poisson like, centered around  $\langle k \rangle$.

\section{Summary}
\label{IV}

In this paper we have generalized the scenario for co-evolution of Axelrod's cultural dynamics and 
network of social ties that was considered in \cite{vazquez2,centola}, by introducing a tolerance 
parameter $Z$ that controls the strength of network plasticity. Specifically, $Z$ fixes the fraction of 
uncommon cultural features above which an agent breaks its tie with a neighbor (with probability equal to the cultural 
dissimilarity), so that, the lower the $Z$ value, the higher the social network plasticity.

Our results show that the network plasticity, when controlled by the tolerance parameter, has competing 
effects on the formation of a global culture. When tolerance is highest, network plasticity promotes cultural 
globalization for values of the initial cultural diversity where multiculturalism would have been the outcome 
for fixed networks. On the contrary, for intermediate values of the tolerance, the network plasticity produces 
the fragmentation of the (artificial) society into disconnected cultural groups for values of the initial cultural 
diversity where global cultural consensus would have occurred in fixed networks. For very low values of the 
tolerance, social fragmentation occurs during the system evolution, but the network plasticity is so high that 
it allows the final cultural homogenization of the transient groups for very high values of the cultural diversity.
Intermediate tolerances promote multiculturalism, while both extreme intolerance and extreme tolerance 
favor the formation of a global culture, being the former more efficient than the latter. 

\begin{acknowledgments}

This work has been partially supported by MICINN through
Grants FIS2008-01240 and FIS2009-13364-C02-01, and by
Comunidad de Arag\'on (Spain) through a grant to FENOL group. Y. M. was partially supported by the FET-Open project
DYNANETS (grant no. 233847) funded by the European Commission and by
Comunidad de Arag\'on (Spain) through the project FMI22/10. 

\end{acknowledgments}


\begin{thebibliography}{99}
\bibitem{note} The decrease to zero of a positive cultural overlap cannot be strictly excluded, though it may be considered  
as a non typical event.
\end{thebibliography}
\end{document}